# Calibratable Hetero-NodeRank for measuring node influence

Qiwei Ma[1] and Zhaoya Gong[2]*

1. School of Architecture, Tsinghua University, China

2. School of Geography, Earth and Environmental Sciences, University of Birmingham, UK

*Corresponding author

Email: z.gong@bham.ac.uk

# Abstract

Node influence metrics have been applied to many applications, including ranking webpages on internet, or locations on spatial networks. PageRank is a popular and effective algorithm for estimating node influence. However, conventional PageRank method considers neither the heterogeneity of network structures nor additional network information, causing a major impedance to performance improvement and an underestimation of non-hub nodes' importance. As these problems are only partially studied, existing solutions are still not satisfying. This paper addresses the problems by presenting a general PageRank-based model framework, dubbed Hetero-NodeRank, that accounts for heterogeneous network topology and incorporates node attribute information to capture both link- and node-based effects in measuring node influence. Moreover, the framework enables the calibration of the proposed model against empirical data, which transforms the original deductive approach into an inductive one that could be useful for hypothesis testing and causal-effect analysis. As the original unsupervised task becomes a supervised one, optimization methods can be leveraged for model calibrations. Experiments on real data from the national city network of China demonstrated that the proposed model outperforms several widely used algorithms.

# 1  Introduction

Network analysis has been a premier approach in many research fields including transportation pattern, protein-protein interaction, disease spread and so on (Bell & Iida, 1997; Lindquist et al., 2011; Schwikowski et al., 2000). Node influence metrics are important measures that rank or quantify the influence of every node within a graph. Their representative applications include measuring the influence of each person in a social network, the role of nodes or actors in the Internet, transportation networks, or urban networks, and the preading power of a node in disease dynamics on human networks. Various indices including betweenness centrality, eigenvalue centrality, degree, or k-shell are introduced to describe node influence in the past thirty years (Carmi et al., 2007; Newman, 2005; Ruhnau, 2000; J. Zhang & Luo, 2017).

PageRank, originally proposed for web mining and later modified in many other fields, is one of the most widely used algorithm for identifying the node influence metrics via measuring the centrality index of each node. PageRank method is invented to measure the global importance of web pages with the behavior of a random surfer, which is a Markov chain process on web graph (Page et al., 1999). In general, the node influence represented by its PageRank value could be interpreted as the probability of an event occurs on this node, such as the infection risk of an individual in a population, or the overload risk of a power station in the national power grid. The distinct advantage of PageRank lies in two aspects. The first advantage is its perceived effectiveness. Using delicate methods such as power iteration, the ranking value of each node in an extremely large network could be calculated rapidly. Secondly, the mechanism of PageRank is explicit. A node's ranking score are treated as inherent votes from its neighbors though connections, which exploits the philosophy of collective wisdom. The popularity of PageRank has inspired lots of researches to develop domain-specified works, producing hundreds of variants(Jing & Baluja, 2008; Sangers et al., 2019; Wang et al., 2008; Z. Zhang et al., 2017).

A key challenge for both the original PageRank and most of its variants is that they are unable to account for heterogeneous network topology. Network heterogeneity is an essential topological feature that could be found in many real-world large-scale complex networks. In these networks, a few nodes, strongly linked with lots of other nodes, are so called hub nodes, while most other nodes have quite limited links or weak connections (Hu & Wang, 2008a). Studies in many fields have shown that the heterogeneity of a network has significant impact on network performance (Hu & Wang, 2008a; Malik et al., 2019; Shekarappa et al., 2017). However, most PageRank-like algorithms lose key semantic information of network heterogeneity, making them fail to capture the node influence of the vastly more numerous non-hub nodes (Šikić et al., 2013a). In scale-free networks, important events are not always triggered by hub nodes. In contrast, the failures occur on less influential nodes may also cause systematic problem, even the collapse of the whole system. These phenomenons have been witnessed in many fields (Cha et al., 2010; Christensen, 2013; Reperant, 2010).

In addition, incorporating additional network information in addition to network topology has also been proved to boost the performance of node influence estimation. Additional information are mainly node-related features (note attributes) that represent the unique impact of node charactersitics on its influence. For example, although the number of visits to a webpage is highly related with its importance, the quality, richness and up-to-dateness of a webpage's content also have significant impact on the stay time, revisiting and other elements that influence its importance. additional network information have been leveraged in many studies and improved the performance of various tasks on graphs (Cha et al., 2010; Everett & Borgatti, 2012; Jia et al., 2017; Kim & Leskovec, 2011). However, this question has not been studied thoroughly in most of the PageRank-related works.

To address above issues, we present a new PageRank-based model framework, dubbed Hetero-NodeRank (HNR), which accounts for heterogeneous network topology and incorporates node attributes as additional network information in order to capture both link-based and node-based effects in measuring node influence. Specifically, this framework allows the variation of link-based effects on node influence across nodes given their topological structures. It also provides an approach to both combining a range of node attributes to estimate node-based effects on node influence and examinining decomposed effects of individual attribute. Furthermore, this framework enables the calibration of HNR model against empirical data, which transforms the original deductive approach into an inductive one that could be useful for hypothesis testing and causal-effect analysis. As the original unsupervised task becomes a supervised one, optimization methods can be leveraged for model calibrations. Experiments were conducted to compare our model with other models on a city network dataset and the results show that our model can outperform other models.

The rest of the paper is organized as follows. The PageRank algorithm and its extensions are reviewd in Section 2. The design as well as the calibration of our model framework are introduced in Section 3. In Section 4, the performance of our model is compared with other baseline algorithms on a city network dataset and the results are discussed. Conclusions and future works are given in the last section.

# 2 Literature review

## 2.1 Original PageRank and its variants for various applications

In the original PageRank algorithm, for a network $G$ with $N$ nodes, the PageRank value of a node $u$ in a network of $n$ nodes with adjacency matrix $w(u)$ is computed from:

$$PR(u) = (1-d)x(u) + d \sum_{v \in G} w_{uv}^T \cdot PR(v) \tag{1}$$

Where $PR(u)$ and $PR(v)$ represents the ranking value of node $u$ and a neighborhood node $v$ respectively. $d$ is the global damping factor referring to the probability of a user following direct links for web browing, solving the so called rank sink problem; othervise, a user jumps randomly to a unlinked webpage with a probability $1 - d$. Conventionally, $d$ is set to 0.85 (Brin & Page, 1998). Apparently, $1 - d$ represents PageRank distribution from non-directly linked pages. $x(u)$, dubbed teleportation vector, is the probability distribution of all webpages for a user to arrive once he/she decides to jump to a unlinkded webpage. Most previous studies adopt a uniform distribution for $x(u) = 1/N$. $w_{uv}^T$ denotes the transpose of the adjacency matrix of $u$. The rank score $PR(v)$ of a node is evenly divided among its outgoing links, which means $w_{uv} = 1$ if there is a direct link between node $i$ and node $j$. Otherwise, $w_{uv} = 0$.

Since the introduction of PageRank, various variants have been proposed to suit different domains and research goals. Among the variations, the first type and the most strenghtforward improvement are to find the more suitable value for the damping factor. Specifically, there are two popular values stand out in the literature for web search: α = 0.5 (Avrachenkov et al., 2007; P. Chen et al., 2007) and α = 0.85 (Page et al., 1999; Najork et al., 2007). These values have been tested through mathematical analysis and empirical explorations. Meanwhile, most of them are domain-specified and are faraway from being generalizable.

The second type of variants focus on exploiting and leveraging network topology information. In order to capture further information aside from directly linked nodes, N-step PageRank(L. Zhang et al., 2007) replacing the original transition matrix with one whose entries are based on the number of a node's N-step neighbors. To weight nodes with attractiveness, Xing and Ghorbani(2004) proposed a Weighted PageRank (WPR) algorithm that makes use of the in-degree of nodes to determine their attractiveness and computes ranking scores of nodes proportional to their relative attractiveness values. Consequently, a node with a low attractiveness value will be ranked lower than a node with a higher attractiveness value. WPR has been extended to incorporate different network attributes to estimate the centrality of nodes under various contexts of applications. Some studies identified the attractiveness of locations using attributes of population flows as weights. In the PlaceRank algorithm proposed by El-Geneidy and Levinson (2011), the cumulative job opportunities of locations is measured by incorporating the amount of worker flow. Similarly, Zhong and Liu (2011) employ proportions of migrations together with geographic distances between origin and destination cities to measure attractiveness scores of cities for permanent population.

The third type of variants aims at exploring external observable data in addition to the topological information of networks to enhance the efficacy of the centrality estimation. By introducing observable data, it becomes possible to modify the Markov chain transition probabilities distribution in light of the observed transitions. A most common approach is manual tagging. For example, TrustRank (Gyongyi et al., 2004) utilized human evaluation data (spam vs. non-spam) as well as initial ranking scores valided by human experts for a sub-set of nodes (instead of random scores). While the former one acts as training data, the later one is combined with a PageRank-based method to produce the final rankings. Spatial data can also be instrumental in informing spatial networks for centrality modeling. Based on the understanding of geographic process (Wen, 2015), two geographically modified PageRank algorithms, Distance-decay PageRank (DPR) and Geographical PageRank (GPR), incorporating geographic considerations into PageRank algorithms were proposed to identify the spatial concentration of human movements in geospatial networks. DPR uses geographic distance to weight the links between locations, while GPR adpots a spatial interaction function that incoporates in-degree as attractiveness and geographic distance as impedance. For page ranking tasks, observed user transition data are the most insightful information to improve network topologies. Berkin et al. (2008) combien this data with user segmentation to create a 'user-sensitive' PageRank, while the same effort is done by researchers at Microsoft (Y. Liu et al., 2008), who use both observed user transitions and holding times to generate a continuous-time Markov chain and create a so-called BrowseRank algorithm.

## 2.2  Heterogeneous networks and node influence

Research in past decades has advanced the methods to identify most influential nodes as a result of almost any spreading process on a given network via developing various centrality measures, such as betweenness (Freeman, 1978; Friedkin, 1991), eigenvalue (Bonacich, 1987), degree (Albert & Barabási, 2002), or k-shell (Kitsak et al., 2010). However, they are less informative for the vast majority of nodes that are not very influential (Lawyer, 2015). The seminal work by Borgatti and Everett (2006) demonstrated that accuracy of these centrality measures is highly dependent on network topology while complex networks are commonly featured with heterogenous topology. Heterogeneous networks often refer to networks with diverse topological characteristics, such as linkage patterns based on degree distribution and related network statistics. Network heterogeneity has been explored in various studies (Horvath & Dong, 2008; Hu & Wang, 2008b; Stam & Reijneveld, 2007).

Recent studies provide more empirical evidence that traditional centrality measures can considerably underestimate the role of non-hub nodes in spreading processes and some of them, e.g., PageRank and k-shell, are highly sensitive to pertubations in network topology (Šikić et al., 2013; Ghoshal & Barabsi, 2011; Adiga et al., 2013). In addition, these centrality measures tend to be less appropriate for the vast majority of nodes that are not very influential, as they are not designed to quantify how influential a node is under the interplay between dynamic spreading processes and network topology (Bauer & Lizier, 2012; Estrada & Rodriguez-Velazquez, 2005). Various node influence metrics have been proposed to remedy the deficiency of centrality measures (Travençolo & da F. Costa, 2008; Klemm et al., 2012; Bauer & Lizier, 2012; Bauer & Lizier, 2012; Lawyer, 2015). For example, accessibility (Travençolo & da F. Costa, 2008) uses the diversity of random walks of a fixed length to measure how reachable the rest of the network is from a given start node, while explicitly accouting for transmission rates of a dynamic spreading process. Moreover, expected force (Lawyer, 2015) focuses on the local network topology of heterogeneous networks to derive the expected values of transmission force by summarizing the distribution of transmission potentials after a small number of transmission events arising from a seed node.

Along the same line, efforts has been made to the PageRank framework to consider the local topology of heterogeneous networks by allowing for a varying local parameterization instead of using global parameters. Given that user preference may lead to different browsing behaviors (Huberman et al., 1998; White & Drucker, 2007), it was natural to generalize the damping factor $d$ as determined by a random variable following a probability distribution to reflect users' browsing preferences (Constantine & Gleich, 2009), while the uniform value for the damping factor in original PageRank is the mean of the random variable. In this work, the general damping factor approach was demonstrated to outperform the original one by accouting for the heterogeneity of user preference. Becasue for most real-world networks following a heavy-tailed distribution, a mean oversimplifies the heterogeneity in the distribution and inevitably weights more on the hub nodes while tends to underestimate non-hub nodes. In a family of link-based ranking algorithms propagating page importance through links, Baeza-Yates et al. (2006) generalized the damping factor as a function of path distance, such that a direct link carries more weights than a link through a long path, capturing the neighborhood topology. However, nearly all these works assumed predefined distributions or functions to generalize the damping factor, which is not only theoretically limited but also yet to be tested against real-world data for their empirical validity.

## 2.3  Node attribute as an additional source for node influence

As discussed in many recent studies (Gomez-Rodriguez et al., 2012), node influence of a network comes from two major sources: network topology and other sources. Taking websites as an example, the influence of a website not only depends on how many other websites links to it, but also on its own quality and themes/topics of this website. Node attribute, as one of the most important sources rather than network topology formed by links, has been investigated for its effects on node influence. Saito, Kazumi, et al. (2011) incorporated node attribute dependency to derive diffusion probability and time-delay parameters on links, while addressing the deficiency of uniform parameter value assumption that cannot be justified for heterogeneous networks. In another work conducted by Zhang (2016), the relationship between node attributes and node centrality was explored using a stepwise linear regression method. The result also shows that node attribute can significantly affect the node influence.

Within the PageRank framework, many researches tried to leverage node attributes to construct or extend non-uniform teleportation vectors to better capture node influence. For unsupervised estimation methods, node attributes could directly improve the estimation accuracy by enriching network features. For example, Haveliwala (2003) proposed a context-sensitive ranking algorithm called topic-sensitive PageRank for web searching. Using representative topics of webpages to inform the construction of non-uniform teleportation vectors, it can produce different sets of rankings biased for specific topics and thus improves the search results. Furthermore, in order to measure user influence on Twitter, Topic-specific TwitterRank (Weng et al., 2010) was developed by using the number of tweets users published and the distribution of user interests on specific topics to construct topic-dependent user relationship networks and teleportation vectors. The topic-specific rankings for a range of topics were then combined to evaluate the overall influence of Twitter users, which outperforms PageRank and topic-sensitive PageRank methods as it accounts for both the topical similarity between users and the link structure. Taking a more general approach, AttriRank (Hsu et al., 2017) combined the base network with a constructed similarity network for node attribute to jointly determine the ranking scores for nodes, under the assumptions that nodes with similar attributes should receive similar ranking scores despite linkage matters as well.

Many semi-supervised modifications tend to treat prior knowledges as constrains, making the estimation process an optimization task (Y.-L. Chen & Chen, 2011). For instance, Adaptive PageRank (Tsoi et al., 2003) rearranges the ranking from a viewpoint of administrators. It uses the ranking (the order of the pages) as a priori constraint, thus the main PageRank equation becomes an objective function that minimizes the Euclidean norm between the ranking score of their model and predefined target values scored by experts. Semi-Supervised PageRank (Gao et al., 2011) proposed a semi-supervised approach to a more accurate estimation of the damping factor. It includes node features and edge weights separately to improve the reset probability matrix and the transition matrix of the original PageRank.

During the computation process, the optimal parameters and ranking scores are obtained simultaneously by minimizing the objective function subject to the constraints.

# 3 Methodology

We propose a new PageRank-based model framework, dubbed Hetero-NodeRank (HNR), which captures both link- and node-based effects in measuring node influence by accounting for heterogeneous network topology and incorporating node attributes.

## 3.1 Local damping factor for heterogeneous networks

In this paper, we consider the most general condition of a weighted and directed network $G = \{V, E, W\}$, where $V = \{1, \ldots, n\}$ is the set of nodes, $|V| = N$, $E \subseteq V \times V$ is the set of edges, and $W = \{W_{ij}\}$ is the weighted adjacency matrix $W = |w_{ij}|_{N \times N}$, where $W$ is a standardizing matrix for each column in the OD-matrix.

To address the local variation of the damping factor, the simple global damping factor $d$ of original PageRank algorithm is expanded to a $N$ dimension vector $D$. There are $m$ different damping factors in $D$ with $0 < m \leq N$. The ranking value $PR(u)$ of node $u$ is calculated as follows.

$$PR(u) = (1 - d(u))\frac{1}{N} + d(u) \sum_{v \in G(N)} w_u^T \cdot PR(v) \tag{2}$$

where $d(u) \in D$ $and$ $0 \leq d(u) \leq 1$ is the locally defined damping factor for node $u$; it mediates the link-related effect on a node. $(1 - d(u)) \frac{1}{N}$ is the contribution of the node-related effect to node influence. Note that $d(u)$ can be viewed as a weighting scheme reflecting the tradeoff between the link-related and node-related effects on the calculation of node influence. $d(u)$ as the local damping factor plays a key role in capturing the heterogeneity of link-related effects. When $m = N$, each node has a distinct damping factor that models its individual behaviors, and the heterogeneity of the network is fully captured. On the other side, it is rarely possible to collect sufficient data for each node, which usually leads to severe overfitting problems. Thus, we allow $m \leq N$ to provide a group-based parameter scheme to provide an operatable data-driven approach to approximate the complicated pattern of internal effects while keeping the generality of the model. We choose an optimal parameter for each group, which represent the local impact of link-related effects at an acceptable precision. Meanwhile, data from each cluster will provide enough samples for parameter estimation.

## 3.2 Incorporating node attribute in node influence

In the heterogeneous network $G$, node attribute matrices of the whole network could be represented by $X_G = \{X_1, \ldots, X_n\}$, where $X_i = \{x_1, \ldots, x_m\}$ is an $m$-dimension standardized attribute vector of a single node $i$. Then some kind of combination function $f(X_i)$ is applied to combine these attributes into a single node-based effect. In this study, we simply use a linear combination function $L(X_i)$ , but it could be easily substituted with other functions such as weighted SoftMax function[1] and so on.

---

[1] https://en.wikipedia.org/wiki/Softmax_function

$$L(X(i)) = \sum_{j=1}^{m} a_j x_j \qquad (3)$$

$A(i) = \{a_1, \dots, a_m\}$ is the m-dimension weight vector that represents the importance of each attribute. The original PageRank algorithm is then expended to:

$$PR(u) = (1-d) \sum_{i=1}^{k} a_i x(u)_i + d \sum_{v \in G(N)} w(u)^T PR(v) \qquad (4)$$

For a network without any node attribute, a uniform damping vector is then adopted where $x_i = 1/N$ for $i = 1$ to $N$, and equation (4) is simplified to formula 1.

## 3.3  Combined model

Decomposing node attributes brings us two main benefits. First of all, it boosts the ranking performance. As is proved in many studies, node attributes can complement the network structure, making the estimated node influence matrices more precise. Secondly, incorporating node attributes strengthen the power of interpretation of PageRank. Similar to linear regression, each element $a_i$ in $A(i)$ act as the coefficient of a node feature $x(u)_i$. Therefore, we not only obtain the contributions of node- and link-based effects by estimating $d$, but also explore the contribution of each single node attribute through $A(i)$. Also, statistical diagnostics such as significance level, confidence interval and so on could be computed through simulation techniques such as bootstrapping sampling.

Finally, by combining the two methods we discussed in 3.1 and 3.2, we propose a flexible framework where both damping factors $d(i)$ and the coefficients of node-based effects $A_i$ are allowed to vary simultaneously to fully capture the heterogeneity of the network.

$$PR(u) = (1 - d(u)) \sum_{i=1}^{k} a_i x(u)_i + d(u) \sum_{v \in G(N)} w(u)^T PR(v) \qquad (5)$$

As is described in equation (5), the influence of a node in a heterogeneous network $G$ could be divided into two parts. The internal part $\sum_{v \in G(N)} w(u)^T PR(v)$ aggregates the information from its neighbors according to the 1st-order Markov assumption. Then its contribution on the node influence is weighted by the locally defined damping factor $d(i)$. On the other side, the impact level of the node-based effect is defined by the linear combination $\sum_{i=1}^{k} a_i x(u)_i$, with its contribution weighted by $1 - d(i)$.

## 3.4  Empirical estimation of model parameters

Node influence matrices are latent values that cannot be measured directly, thus original PageRank algorithm and most of its variants adopt the unsupervised learning method and empirically defined parameters to estimate them. Thanks to the emergence of big data, it becomes more and more possible to exploit various label data that is potentially correlated to latent node influences for approximation. These indicators includes economic growth data(Barcelos et al., 2017; Takes & Heemskerk, 2016), traffic monitoring data(Kazerani & Winter, 2009), observed vitality data(Yue & Zhu, 2019) and so on. Therefore, it is feasible to convert the traditional unsupervised tasks to supervised tasks, which means that the model can be calibrated using existing data then generalized to estimate unknown node influence.

Similar to the original PageRank algorithm, node influence matrices in our method are calculated recursively. Therefore, there is no close form solutions for solving model parameters. Most previous works focused on two approaches. 1) Using fixed parameters or some sort of parameter distributions empirically. 2) Using preliminary parameter searching strategies such as parameter sweeping. However, these approaches do not guarantee near optimal parameter selection. In our framework, due to the introduction of additional parameters including locally defined damping factors and coefficients of node attributes, the parameter space is quite high-dimensional, making the search of parameters more difficult.

To solve the above problem, evolution strategy optimization techniques such as Genetic Algorithms (GA) (Sivanandam & Deepa, 2008), Differential Evolution (DE) (Price, 2013) and Particle Swarm Optimization (PSO) (Kennedy & Eberhart, 1995) are introduced to execute effective search for optimal solutions in high-dimension parameter spaces. The main reason we choose these algorithms is that they use objective function information without the requirement of any gradient information. Thus, it is capable of solving non-convex optimization problem in our framework. Also, they provide means to search irregular space, which is suitable for the model calibration. In this paper, we choose GA as an example to demonstrate the computation process because of its simplicity of interpretation.

GA is a family of computational models based on principles of evolution and natural selection. These algorithms convert the problem in a specific domain into a model by using a chromosome-like data structure and evolve the chromosomes using selection, crossover, and mutation operators (Sinclair, Pierce, and Matzner 1999; see also Whitley, 1994). GA has been developed for decades, and its theory and implement can be found in lots of papers and software. Figure 1 shows the structure of a simple genetic algorithm. It starts with a randomly generated population, evolves through selection, crossover, and mutation. Finally, the best individual (chromosome) is picked out as the final result once the optimization criterion is met (Pohlheim, 2001)[2]. The detailed process is depicked in Figure 1.

---

[2] Using Genetic Algorithm for Network Intrusion Detection

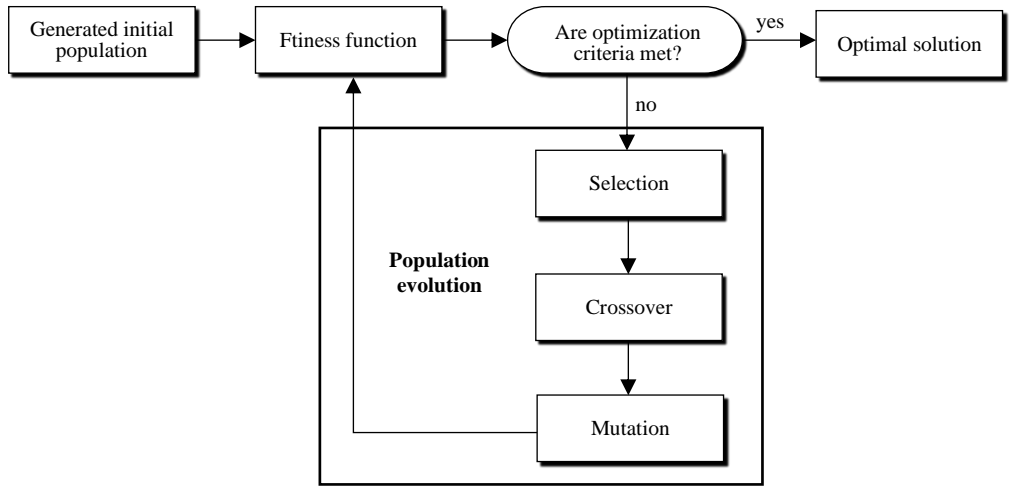

Figure 1. Structure of a simple genetic algorithm (Pohlheim, 2001).

Aside from the general setting of GA models, there are two key points worth further discussing.

**(1) Fitness functions transformation**

The goal of our method is to estimate node influence matrices as close as possible to the ground truth. We denote $\mathcal{F}(\hat{y}, y)$ as the loss function which measures the difference between estimated node influence matrices $\hat{y}$ and the ground truth $y$. The more accurate our model is, the closer $\hat{y}$ is to $y$, and the smaller $\mathcal{F}(\hat{y}, y)$ becomes. Therefore, $\mathcal{F}(\hat{y}, y)$ is naturally defined as a minimization problem. According to specific problems we are solving, lots of loss functions including L1 absolute loss, L2 squared loss, or cross entropy loss and so on could be chosen.

On the other side, as GAs mimic the survival-of-the-fittest principle of nature to make a search process, they are naturally designed for solving maximization problems, which does not suit our problem directly. For minimization problems, to generate non-negative values in all the cases and to reflect the relative fitness of individual, it is necessary to converts our minimization problem $\mathcal{F}(\hat{y}, y)$ to an equivalent maximization problem. A number of such transformations have been developed. For example, a simple fitness mapping can be defined as follows:

$$F(\hat{y}, y) = \frac{1}{1 + \mathcal{F}(\hat{y}, y)} \quad (6)$$

## (2) Population initialization

Population initialization is the first step in the GA Process. Population is a subset of solutions in the current generation. Population $P$ can also be defined as a set of chromosomes. The initial population $P(0)$, which is the first generation is usually created randomly. In an iterative process, populations $P(t)$ at generation $t, t = (0,1,\dots)$ are constituted (Fig.2). Suppose the network is partitioned into $K$ groups, and the $k$th group of the $i$th chromosome have the parameter set $P(t)_i^k = \{d(k), a(k)_1, \dots, a(k)_m\}$. To form the chromosomes of $P(t)$, concatenation operator is applied and $P(t)_i = concat(P(t)_i^k)$. Following the Markov assumption, each element in $P(t)_i$ have the constraint $d(k) \in [0,1]$ and $a(k)_m \in [0,1]$.

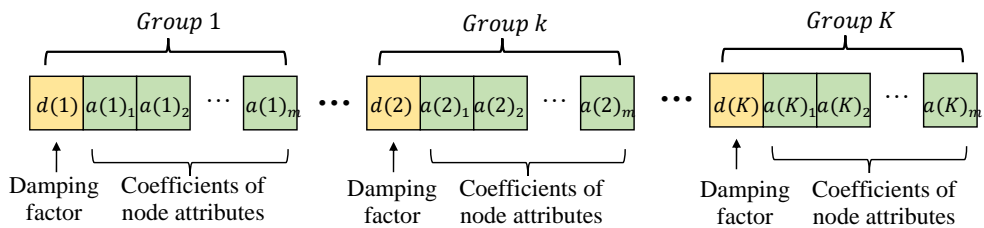

Figure 2. Structure of a chromosome in population P(t).

Incorporating the improved PageRank algorithm into GA models, near optimal solutions could be derived in a data-driven manner. The calibration process begins with the initialization of the population according to the predesigned structure. Then each chromosome is inputted into the model separately to compute its fitness value. Then population evolution operators are executed recursively until the termination condition is met. The detailed process is shown in Fig.3.

**Stage 1: Initialization**

In the initialization stage, a network $G$ is constructed given the OD flow dataset. A proper kind of rule is utilized to divide the network into $K$ groups. The rule could be certain kind of network partition technique, expert knowledge and so on. Then a set of population $P(0)$ that acts as candidates of model parameters is randomly generated. Each individual of the initial population is assigned to the network to represent an initial possible solution of our model. The network topology itself is also treated as another input resource.

A number of ways in which networks can be clustered have been proposed in recent years. For examples, clustering by score, clustering by rank, clustering by rank with fixed cluster dimensions or clustering by rank with variable cluster dimensions using a set regime are all considerable solutions. Concerned by the long-tailed distribution of node influence in most real-world networks, clustering techniques

**Stage 2: Equilibrium process of each individual**

After the configuration of the model, iteration process is carried out to find the ranking value of each node. During each iteration, internal effects are derived by aggregating information from the current ranking values of neighborhood nodes. Meanwhile, node-based effect is calculated using the linear combination of node attribute vector simultaneously. After the ranking value $PR(u)$ is derived by summing both the effects from network topology and node attributes, the equilibrium checking is carried out. If all nodes' current scores are equivalent to their previous ones, a state of equilibrium has been reached, and the current ranking values are regarded as the final *PR* scores. If not, a new iteration will be started. The process will be executed recursively until convergence.

**Stage 3: Evaluation and Evolution**

The output ranking values from stage 2 as well as labels derived from other resources are imported together into a designed loss function $L(\hat{y}, y)$ to evaluate the performance of all sets of parameters by measuring the similarity between prediction $\hat{y}$ and the label values $y$.

If the minimum of the costs $L(\hat{y}, y)$ is below a given threshold, or the max iteration count is reached, the convergence condition is met, and the individual with lowest cost value in the whole population is selected as the final champion, or optimal solution. If not, genetic operators including selection, crossover and mutation are applied to produce a new set of population. The above process is carried out repeatedly until the optimal solution is found.

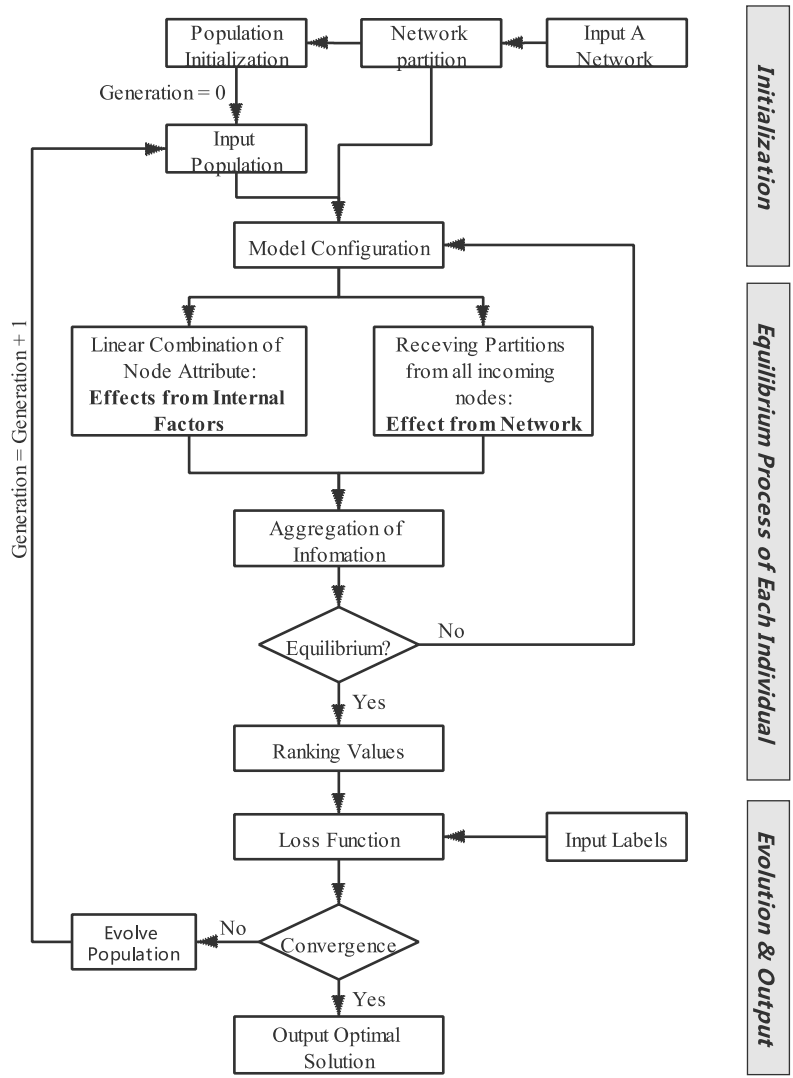

Figure 3. The framework of model calibration.

# 4 Experiments

## 4.1 Datasets, evaluation and baseline models

To verify our proposed model, an experiment is carried out on the national city network of

China mainland (NCN). A small proportion of node samples are used for the calibration of the model, then the node influence matrices of other nodes are estimated. Finally, the performance of our model is compared with other models.

The NCN dataset is composed of 3 parts. *Node and node attributes*. There are 272 nodes in the network, corresponding to 272 cities with available data sources, which is around 70% of all 334 cities in total in China mainland. Also, there are about 62 cities excluded from the dataset for lacking necessary features. The node attribute vector of each city contains 6 feature categories, representing 6 different factors that affect the node influence of this city (Tab. 1). All these featured have been proved to affect the node influence of cities to some degrees (Glaeser & Shapiro, 2001; Da Mata et al., 2007; Sridhar, 2010; T. Liu & Cao, 2011; Kelley & Burley, 2014). *Network links*. Connections between pairs of cities are represented by the short-term intercity travel flow. They form the topology of the network and represent the link-related effect in our model. The data is provided by Baidu Maps, and is collect between 2017/4/1 and 2017/4/7. The direction of travel determines the direction of the connection, while the total number of travelers is the weight of the connection. *Indicator of node influence matrices*. We utilize the growth ratio of GDP from 2017 to 2018 as an indicator of the latent node influences, following previous works(Bird, 2013; Frick & Rodríguez-Pose, 2016). The ratio is calculated by $(GDP_{2018} - GDP_{2017})/GDP_{2017}$. The data is also provided by the 2017 China City Statistics Yearbook. Cities with higher growth ratio are supposed to be more important in the city network, thus is more influential.

Table 1. The description of 6 categories of node attribute

| Data name | Acquired time | Data source | Definition | Effect description |
| --- | --- | --- | --- | --- |
| Persistent Population | 2017/8/31 | Baidu Maps | a persistent resident is a person that lives in the city for more than 6 months(National Bureau of Statistics, 2014) | A city with larger population is more likely to influence other cities |
| GRP | 2017 | 2017 China City Statistics Yearbook(National Bureau of Statistics, 2018) | the total monetary or market value of all the finished goods and services produced | As a broad measure of overall domestic production, it functions as a comprehensive scorecard of economic health for a city |
| The proportion of the second industry | 2017 | 2017 China City Statistics Yearbook | the proportion of the value produced by the secondary sector to the total GRP of a city | This variable indicates how deeply the development of a city relies on manufacturing approximately |
| Accessibility of high-speed railway(hsr) | 2017 | Baidu Maps | The variable is binary. If a city owns one or more hsr stations, the value is set to 1, otherwise 0 | The accessibility of hsr and airport affects the intensity of connections of a city to the whole city network, which |

| Accessibility of airport | 2017 | Baidu Maps | This variable is also binary. A city with one or more airport got value of 1, otherwise 0. | improves its competitiveness |
| --- | --- | --- | --- | --- |
| The yearly mean temperature | monthly ground temperature from 1970-2000 | WorldClim website(Fick & Hijmans, 2017) | mean temperature value averaged by month | The climate situation is related with the economic and social vitality of a city |

For the city network, since the ground truths are real values, we follow the experiments of [3] to use Spearman's rank correlation coefficient as the evaluation metric. To make sure the sampling process is balanced, cross-validation techniques are employed. Each time 30% nodes are randomly sampled for model calibration. Then the ranking values of the rest 70% nodes are estimated using the calibrated model and the fitness score is derived by calculating the Spearman's correlation coefficient between ranking values and GDP growth ratios of these cities. The process is executed 10 times, the final evaluation metric is computed taking the average value of all 10 results. To evaluate model performance on more influential nodes and less influential nodes separately, a head/tail breaks technique (Jiang, 2013) is used according to the heavy-tailed distribution characteristics of NCN dataset. Head/tail breaks is a simple but powerful algorithm for deriving inherent classes or hierarchical levels (Basofi et al., 2015; Jiang, 2016; P. Liu et al., 2020). By evaluating model performance in each head/tail part (Fig. 4), we further investigate the effect of HNR in equally estimating influence of all nodes.

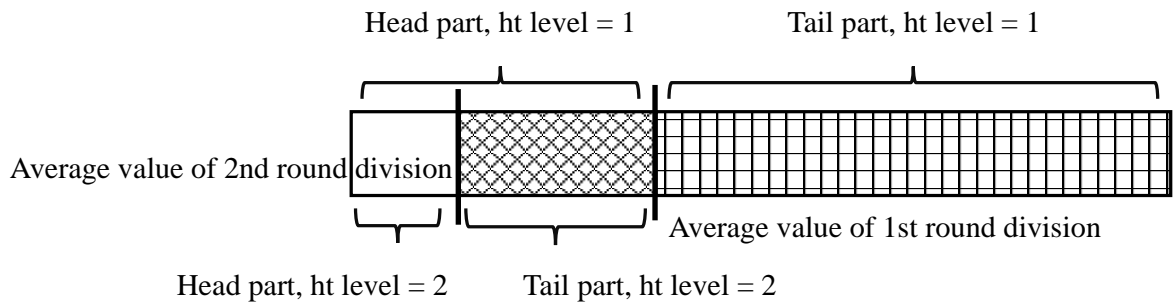

Figure 4. Head/ tail breaks divides values around an average into a few large ones in the head (those above the average) and many small ones (those below the average) in the tail, and recursively continues the division process for the large ones (or the head) until the end condition is met. A term called ht level is introduced to indicate the count of division times when a head/tail part is generated. For example, the ht level of head/tail parts derived through the second division is 2.

We test three baseline models compared with our model: PageRank, Weighted PageRank (WPR) and AttriRank. HNR with parameters partly fixed are also adopted to test link-related effects and node-related effects separately.

**PageRank:** It is our baseline solution to node importance ranking. The only parameter in PageRank is the damping factor, which is heuristically set to 0.85 in our experiment.

**WPR:** Utilizing the in- degree $I_u$ and out-degree $O_u$ information of a node $u$, WPR determines the edge weights to improve ranking reliability. The update rule of WPR is as follows:

$$PR^{(t+1)} = (1-d)\frac{1}{N}1 + dWPR^{(t)}$$

$$W_{uv} = \frac{1}{z_v}\frac{I_u}{\sum_{k \in F_v} I_k}\frac{O_u}{\sum_{k \in F_v} O_k}$$

(7)

where $F_v$ is the set of nodes pointed to by node $v$, and $z_v$ denotes the normalization term for column $v$. $d$ is also set to 0.85 as is adopted in most previous studies.

**AttriRank:** AttriRank adopts a scenario where a random walker simultaneously moves in two graphs G and H. H is a fully connected undirected graph sharing the same node set V of G. Each edge weight between nodes (i,j) in H represents the similarity$ij$ >0 between the corresponding node attributes (xi,xj). Like PageRank, AttriRank can be interpreted by a Markov chain model with the following update rule:

$$PR^{(t+1)} = (1-d)QPR^{(t)} + dWPR^{(t)}$$

$$P_{uv} \equiv \begin{cases} \dfrac{1}{\delta_j} & \text{if directed edge}(v,u) \in E \\ \dfrac{1}{N} & \text{if } \delta_j = 0 \\ 0 & \text{otherwise} \end{cases}$$

(8)

$$Q_{uv} \equiv \frac{s_{uv}}{\sum_{k \in V} s_{kv}}$$

where $r$ $W \in \mathbb{R}^{N \times N}$ and $Q \in \mathbb{R}^{N \times N}$ denote the corresponding transition matrices for graph G and H respectively. $\delta_j$ is the out-degree of node j. Parameter $d \in (0,1)$ controls the random-walk preference ratio between graph G and H. $s_{uv}$ represents the scale of similarity between the attributes of two adjacent nodes $u$ and $v$. AttriRank select Radial Basis Function (RBF) kernel as the similarity definition: $s_{uv} \equiv e^{-\gamma \|x_u - x_v\|_2^2}$, where positive parameter $\gamma$ controls the influence of attribute distances. In our experiment, d follows $Beta(\alpha = 2, \beta = 3)$, and

RBF kernel parameter is set to $\gamma = \frac{1}{K}$ where $K$ is the dimension of the attribute vector, as is suggested by the authors.

**ExF:** ExF is formally defined as follows. Consider a network with one infected node i and all remaining nodes susceptible. Enumerate all J possible clusters of infected nodes after x transmission events, assuming no recovery. After all transmissions without recovery, the force of infection (FoI) of a spreading process seeded from node i is a discrete random variable taking a value in $d_1, \dots, d_J$, allowing for the proportionality constant equal to the transmission rate of the process. Then

$$ExF(i) = -\sum_{j=1}^{J} \overline{d_j}\, log(\overline{d_j}) \tag{9}$$

Where $i$ refers to the seed node and $\overline{d_k} = \frac{d_k}{\sum_J d_j}, \forall k \in J$.

**Partly fixed HNR:** Additionally, we also test our model whose parameters are partly fixed: HNR with uniform node attribute (Equation 2) and HNR with uniform damping factor (Equation 4). The former tests the link-based effects by using local damping factor, while the later tests the node-based effects, respectively. Similar to the full model, these two models are first calibrated using 30% labels of node influence, then their performances are evaluated based on the left 70% dataset.

## 4.2 Results and discussions

We verify the following hypotheses about HNR in this section.

**H1: Does HNR outperform baseline models?**

HNR significantly outperforms other competitors with NCN dataset (Table 2). Surprisingly, ExF gains the lowest correlation index, indicating that it may not be capable of capturing node influence correctly in densely connected networks with highly complicated network topology and node attributes. While the original PageRank shows a relatively poor performance due to ignoring most additional information aside from basic network topology, WPR and AttriRank show plausible improvement with correlation scores of 0.29 and 0.25. Using information from network topology and node-related features separately, HNR(l) and HNR(e) perform better with correlation coefficients of 0.32 and 0.43. The performance is further improved to 0.49 when two types of effects are combined together.

Table 2. Model performance comparison (Spearman's correlation index). (e) denotes using node attributes; (l) denotes using local damping; (el) denotes using both node attributes and local damping.

| PageRank | WPR | AttriRank | ExF | HNR(e) | HNR(l) | HNR(el) |
|----------|-----|-----------|-----|--------|--------|---------|
| 0.12*** | 0.29*** | 0.25*** | -0.10. | 0.43*** | 0.32*** | **0.49*** |

**H2: Is local parameters helping better capturing network topology?**

As is shown in Table 2, HNR (l) gains a correlation score of 0.32, which is higher than the scores of PageRank, WPR and AttriRank, implying that incorporating local parameters substantially captures the heterogeneity of network topology.

**H3: Is incorporating node-related information boosting the performance of the model?**

On the other hand, results using node-related attributes alone are even more close to the ground truth, which demonstrates the importance of node attributes. Comparing to AttriRank which models attributes with pairwise attribute similarities in an unsupervised way, HNR(e) performs much better, proving that using a small portion of sample data to calibrate the model is a superior strategy.

**H4: Does our model helps to lessen the biased estimation of node influence?**

As most PageRank-like algorithms are criticized for emphasizing influential nodes while underestimating less influential nodes, our model could largely lessen the biased node influence distribution.

(1) Our model helps to alleviate the overestimation of more influential nodes. As is shown in figure 5a, as ht level grows, there are less influential nodes included in the head part, and the performance of nearly all algorithms are decreasing, indicating that variation is larger in more influential nodes, making it more difficult to capture the true distribution of node influence using global parameter setting without node attributes. During this process, HNR keeps a relatively high correlation score compared with other competitors. As low correlation score may indicate severe overestimation of influential nodes, our model moderately optimizes the ranking mechanism by setting separated parameters for each ht part, resulting in higher correlation scores in each ht level.

(2) Our model better captures the node influence matrices of less influential nodes. For all tail parts, the performance of each model stays relatively stable (Fig. 5b). Among them, PageRank could poorly capture node influence, as is pointed by previous works. WPR, AttriRank and HNR(l) gain similar performance around the correlation coefficient of 0.3. HNR(e) performs significantly better than all other models except HNR, which again proves the importance of node-related features. Finally, HNR achieve best performance in evaluating less influential nodes.

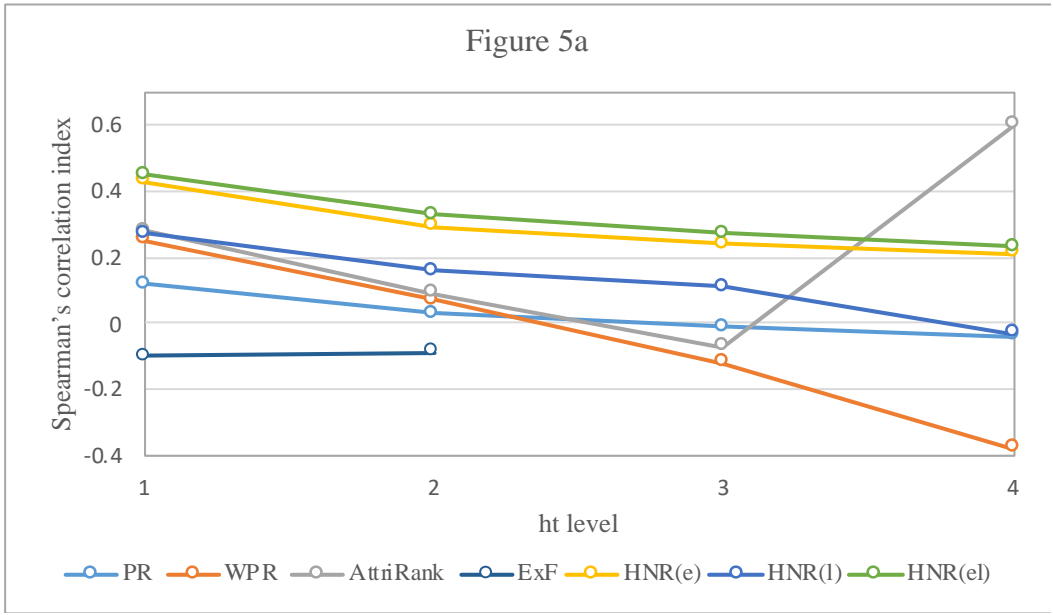

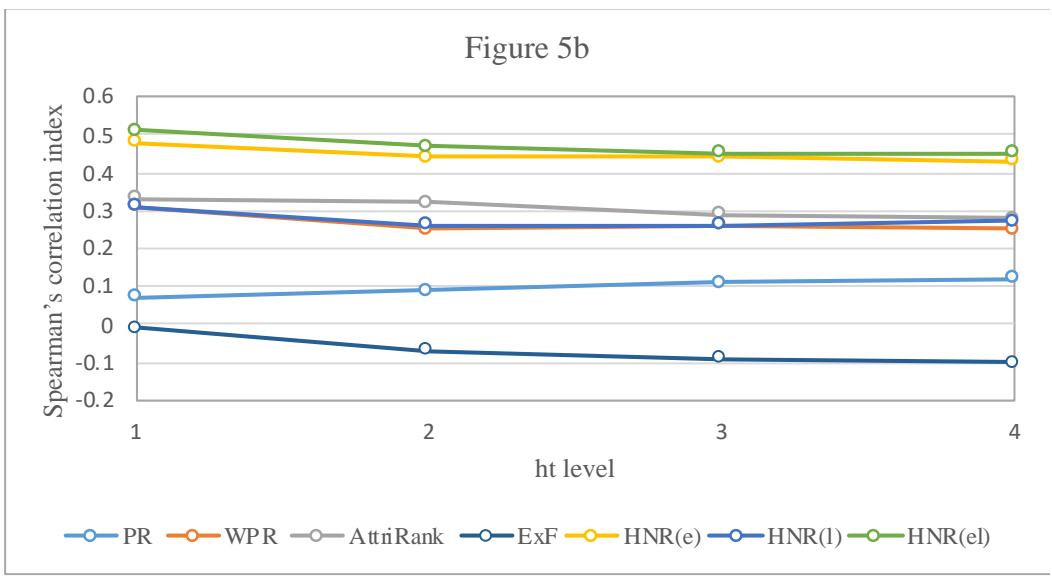

Figure 5. (a) The performance of HNR with compared models on the head part (a) and tail part (b) of the NCN dataset. X axis represents the ht level, and y axis represents the Spearman's correlation index

**H5: Hoe does the size of samples for model calibration affect the performance?**

Figure 6 shows the model performance with different sample size for model calibration. Surprisingly, the correlation coefficient rises rapidly when the proportion of sample data is below 20%. Above this threshold, the growth speed decreases dramatically. When the proportion exceeds 85%, the model performance becomes rather stabilized. This result implies that a near optimal model could be generated using only a small portion of data, which could largely reduce the cost of collecting sample data in practice and ensures the feasibility and generalization ability of our method.

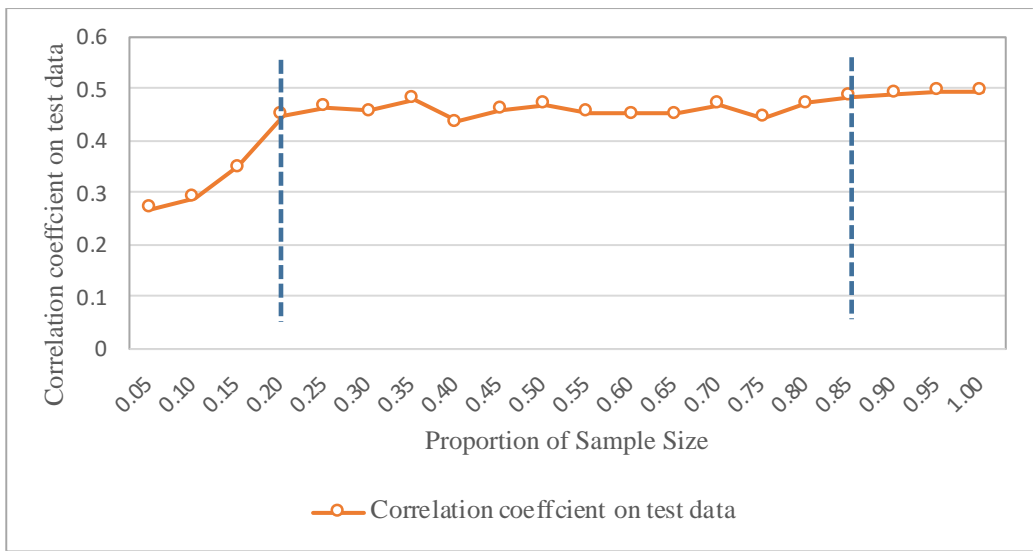

Figure 6. Change of model performance with different proportion of sample size.

# 5   Conclusions and future works

PageRank is still an elastic framework that could gain better ranking performance today. Thus, a general calibratable model framework, Hetero-NodeRank, is proposed in this study to better incorporating the network topology as well as exploiting additional network information. The PagerRank algorithm is extended in three aspects in this paper. First of all, by converting the identical damping factor into a locally defined vector that address the unique behavior of different nodes, the network heterogeneity is better captured, and the influence of both hub nodes and non-hub nodes are estimated in a relatively equaly manner.

Secondly, the original uniform damping vector is also replaced by the combination of various node attributes. By solving the optimal parameter vector, more accurate node influence matrices are derived. Meanwhile, the contribution of each node attribute is revealed together with statistical characteristics, which further provides more evidence on the mechanism of the network dynamics. Thirdly, empirical data is also used as indicators that could be viewed as approximations of node influence. Then the original unsupervised task is shifted to a supervised one, where optimal parameters are calculated using optimization strategies.

Our future works are two-fold. First, advanced optimiazation techniques are considered to be tried to improve the computation efficiency, as evolution optimization approaches are quite computationally complicated. Secondly, more comprehensive experiments will be conducted to investigate the performance as well as the sensitivity of our model to make it more robust.